\documentclass[twocolumn,prl]{revtex4}

\usepackage{graphicx}
\usepackage{enumerate}
\usepackage{siunitx}
\usepackage{braket}
\usepackage{nicefrac}
\usepackage{amsmath}
\usepackage{xcolor}

\begin{document}

\title{Resolved-sideband laser cooling in a Penning trap}

\author{J.F. Goodwin}
\affiliation{Quantum Optics and Laser Science, Blackett Laboratory, Imperial College London, Prince Consort Road, London, SW7 2AZ, United Kingdom.}

\author{G. Stutter}
\affiliation{Quantum Optics and Laser Science, Blackett Laboratory, Imperial College London, Prince Consort Road, London, SW7 2AZ, United Kingdom.}

\author{R.C. Thompson}
\affiliation{Quantum Optics and Laser Science, Blackett Laboratory, Imperial College London, Prince Consort Road, London, SW7 2AZ, United Kingdom.}

\author{(D.M. Segal, Deceased 23 September 2015)}
\affiliation{Quantum Optics and Laser Science, Blackett Laboratory, Imperial College London, Prince Consort Road, London, SW7 2AZ, United Kingdom.}

\begin{abstract}
    We report the laser cooling of a single $^{40}\text{Ca}^+$ ion in a Penning trap to the motional ground state in one dimension. Cooling is performed in the strong binding limit on the 729-nm electric quadrupole $S_{1/2}\leftrightarrow D_{5/2}$ transition, broadened by a quench laser coupling the $D_{5/2}$ and $P_{3/2}$ levels. We find the final ground state occupation to be $\SI{98\pm1}{\%}$. We measure the heating rate of the trap to be very low with $\dot{\bar{n}}\approx\SI{0.3\pm0.2}{\second^{-1}}$ for trap frequencies from \SIrange{150}{400}{\kilo\hertz}, consistent with the large ion-electrode distance.
\end{abstract}

\maketitle
Cold, trapped ions are one of the leading systems with which to study processes that require excellent environmental isolation, including quantum computation~\cite{Nigg2014}, quantum simulation~\cite{Britton2012}, and metrology~\cite{Gill2003,Quint2001}. Penning traps~\cite{Thompson2009,Brown1986} are most widely used for precision measurements on fundamental particles and atomic ions~\cite{Mooser2014,DiSciacca2013,Vogel2015} and have also found applications in quantum information~\cite{Britton2012,Goodwin2015,Porras2006}. Unlike radiofrequency (RF) traps, Penning traps require no oscillating fields, making them suitable for trapping 2- and 3-D Coulomb crystals~\cite{Mavadia2013} and ions in states sensitive to RF perturbation (e.g. Rydberg ions~\cite{Muller2008,Feldker2015}).

Many ion trap experiments require motional ground state confinement, typically achieved via resolved sideband cooling. Sideband cooling was first demonstrated in RF traps many years ago~\cite{Diedrich1989,Roos1999} but, due to the technical complexities associated with Penning traps, had not yet been realised in this system.

Ground state cooling is of particular importance to quantum gates, and recent years have seen the application of these gates to precision measurement~\cite{Chou2010,Shi2013,Wolf2016}. The ability to apply quantum logic spectroscopy to ions in Penning traps will greatly increase the precision of experiments that necessitate their use~\cite{Smorra2015,Cornejo2014}. Coherent control of the motional state also underpins many experiments in quantum thermodynamics~\cite{Poyatos1996,Home2006}, a field where the very low heating rates achievable in Penning traps offer a distinct advantage.

In this Letter we demonstrate the application of resolved optical sideband cooling to the Penning trap, cooling the axial motion of a calcium ion to its quantum ground state with 98\% probability. We demonstrate our ability to coherently manipulate the ion's electronic state by observing its Rabi dynamics. Finally, we measure the ion heating rate, which we find to be the lowest reported in the literature to date for any type of trap~\footnote{This is the first direct measurement of the single-ion heating rate in a Penning trap; a recent measurement by Sawyer \emph{et al}~\cite{Sawyer2014} provided an indirect constraint via the heating rate scaling of large crystals.}. The low heating rate is consistent with the large trap size, which has a characteristic dimension of $d_0=\SI{1.32}{\cm}$, and distance to the nearest electrode $d=\SI{1.08}{\cm}$.

We have previously reported work on resolved sideband spectroscopy of an ion in a Penning trap~\cite{Mavadia2014}. The experiments described here use a modified version of the same apparatus. We trap a $^{40}\text{Ca}^+$ ion in a Penning trap consisting of a stack of cylindrical electrodes, held in a $\SI{1.85}{\tesla}$ axial field provided by a superconducting solenoid magnet. Doppler cooling in the axial and radial directions is performed using two lasers at \SI{397}{\nano\metre}, tuned to two components of the S$_{\nicefrac{1}{2}} \leftrightarrow \textrm{P}_{\nicefrac{1}{2}}$ transition, with four laser frequencies around $\SI{866}{\nano\metre}$ applied along the trap axis to repump population in the D$_{\nicefrac{3}{2}}$ states. At high magnetic fields, $j$-state mixing~\cite{Crick2010} provides a small branching ratio of $4.2\times10^{-7} B^2$/T${^2}$ to the D$_{\nicefrac{5}{2}}$ manifold, necessitating four additional repump laser frequencies around $\SI{854}{\nano\metre}$. A laser at $\SI{729}{\nano\metre}$, addressing the electric quadrupole S$_{\nicefrac{1}{2}} \leftrightarrow \textrm{D}_{\nicefrac{5}{2}}$ transition, is used for resolved sideband spectroscopy via the electron shelving technique. Details of the trap, lasers and spectroscopy scheme can be found in \cite{Mavadia2014}, \cite{MavadiaThesis} and \cite{GoodwinThesis}.

We have made two major changes to the experiment to enable us to perform sideband cooling. The first is an increase in the 729-nm laser power using a tapered amplifier. This increases the power at the ion from $\SI{4}{\milli\watt}$ to $\SI{40}{\milli\watt}$, providing Rabi frequencies of up to $\Omega_0 / 2\pi\approx\SI{50}{\kilo\hertz}$.

The second change is to introduce an oscillating quadrupolar `axialisation' field~\cite{Savard1991} coupling the magnetron and modified cyclotron radial trap modes. Our previous work \cite{Mavadia2014} was performed without any oscillating fields, using a radial cooling beam with an intensity gradient across the trap centre to cool the otherwise-unstable magnetron motion~\cite{Itano1982}. For higher axial frequencies, the intensity gradient necessary for effective cooling increases, and with our current optical system we are unable to reliably cool ions above axial frequencies of $\SI{200}{\kilo\hertz}$. The axialisation technique~\footnote{Note that this technique is sometimes also referred to as `sideband cooling', but is unrelated to the laser cooling method described in this paper.}, which had been used by the group in earlier laser cooling experiments~\cite{Powell2002, Phillips2008, Hendricks2008}, works at all trap frequencies but has the disadvantage of introducing an RF electric field and associated micromotion. Fortunately, the RF potential required for axialisation is several orders of magnitude lower than that used for ponderomotive trapping in Paul traps, and in this experiment never exceeds \SI{50}{\milli\volt}. For a typical trapping potential of \SI{200}{\volt}, the micromotion amplitude is expected to be \SI{<10}{\nano\metre} for an ion situated \SI{10}{\micro\metre} from the RF null.

The ion is Doppler cooled, before one of the two 397-nm lasers is switched off to optically pump population into the S$_{\nicefrac{1}{2}}(m_j=-\frac{1}{2})$ sub-level. For the axial trap frequencies used in this experiment ($\omega/2\pi=\SIrange{150}{400}{\kilo\hertz}$), the ion remains outside the Lamb-Dicke regime ($\eta_{729}^2(2\bar{n}+1)\approx\numrange{8.1}{1.1}$) after Doppler cooling. For the result presented below, at $\omega/2\pi=\SI{389}{\kilo\hertz}$, this does not prevent efficient sideband cooling on the first red sideband. However, approximately 0.17\% of the population is initially in oscillator states higher than $n=150$. As the first red sideband Rabi frequency is very close to zero for this state, any population cooled from higher phonon states will accumulate in a narrow distribution of states immediately above this level. To prevent such Fock-state population-trapping it is necessary to alternate cooling on the first-order and a higher-order sideband, as demonstrated by Poulsen \emph{et al}~\cite{Poulsen2012}.

Sideband cooling is therefore performed by applying the 729-nm laser alternately to the first and second red sidebands of the S$_{\nicefrac{1}{2}}(m_j=-\frac{1}{2})\leftrightarrow\textrm{D}_{\nicefrac{5}{2}}(m_j=-\frac{3}{2})$ transition. The scattering rate is increased by using a weak, 854-nm quench laser to empty the D$_{\nicefrac{5}{2}}$ level via P$_{\nicefrac{3}{2}}$, which rapidly decays at 393-nm to S$_{\nicefrac{1}{2}}$. For small quench laser saturation parameters, the P$_{\nicefrac{3}{2}}$ level can be adiabatically eliminated and the system behaves like a two level system with a linewidth set by the properties of the quench laser and upper level \cite{Marzoli1994}. The effective D$_{\nicefrac{5}{2}}$ linewidth is measured spectroscopically and the quench laser intensity is adjusted to give $\tilde{\Gamma}/2\pi\approx\SI{50}{\kilo\hertz}$. The second 397-nm laser and 866-nm repump lasers are applied continuously to ensure that population decaying on the P$_{\nicefrac{3}{2}}\leftrightarrow\textrm{S}_{\nicefrac{1}{2}}$ transition is optically pumped into the correct ($m_j=-\frac{1}{2}$) ground state sub-level. After sideband cooling, electron shelving spectroscopy is performed as described in~\cite{Mavadia2014}.

In the limit $\Omega_0\ll\omega$, for the cooling cycle used in this experiment, the sideband cooling limit is given by
\begin{equation}\label{eq:sbcLimit}
    \bar{n}_{lim}=\left(\frac{\tilde{\Gamma}}{2\omega}\right)^2\left[\frac{\tilde{\eta}^2+\eta_q^2+\eta_r^2+\tilde{\eta}_r^2}{\eta^2}+\frac{1}{4}\right],
\end{equation}
where $\eta$ and $\tilde{\eta}$ are Lamb-Dicke parameters associated with absorption on the 729-nm transition and emission on the 393-nm transition, $\eta_q^2$ is the Lamb-Dicke parameter associated with absorption at \SI{854}{\nano\metre} from the quench laser, and $\eta_r$ and $\tilde{\eta}_r$ are Lamb-Dicke parameters associated with emission and absorption at \SI{397}{\nano\metre} during optical pumping. Geometric factors due to absorption and emission patterns are included in these Lamb-Dicke parameters.

To achieve the highest probabilities of ground state occupation, especially at lower trap frequencies, we reduce the intensity of the 729-nm laser to 25\% of its maximum during the final stage of cooling, so that $\Omega_{0}/2\pi\approx\SI{25}{\kilo\hertz}$.

\begin{figure}[t]
%\begin{flushleft}
\includegraphics[width=\columnwidth]{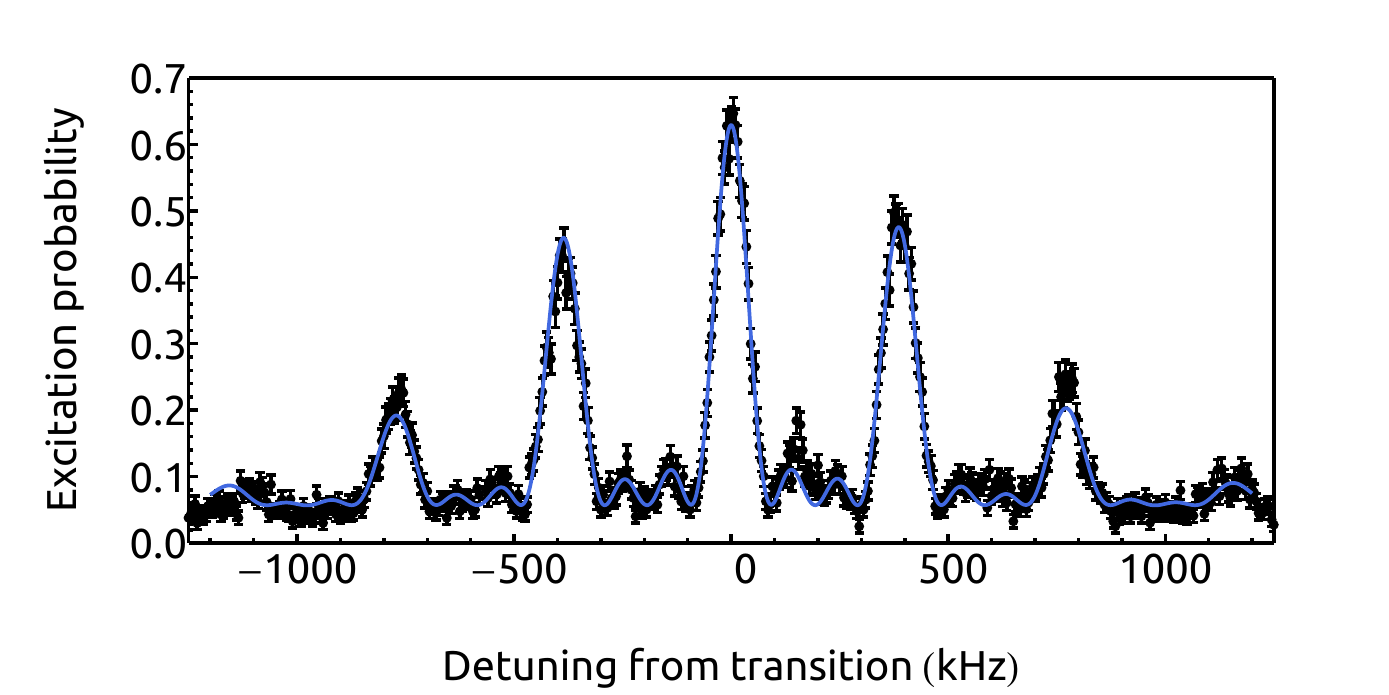}
\caption{(Color online) Typical axial motional sideband spectrum after Doppler cooling, with a \SI{10}{\micro\second} probe pulse length. The solid line is a fit to the Rabi dynamics of a thermally distributed population, giving $\bar{n}=\SI{24\pm1}{}$, equivalent to a temperature of $\SI{0.45\pm0.02}{\milli\kelvin}$.}\label{DopplerFit}
%\end{flushleft}
\end{figure}

The ion is trapped with an axial frequency of $\SI{389}{\kilo\hertz}$ and Doppler cooled for $\SI{5}{\milli\second}$. We measure an axial temperature of $\SI{0.45\pm0.02}{\milli\kelvin}$, consistent with the Doppler cooling limit of $\SI{0.45}{\milli\kelvin}$ and corresponding to an average phonon number of $\bar{n}=\SI{24\pm1}{}$. Figure \ref{DopplerFit} shows a typical axial spectrum after this step, where each data point is the average of 400 repeats. We do not measure a radial temperature during these experiments, but independent measurements suggest this is several times higher than the axial temperature~\cite{Mavadia2014}, typically \SI{3}{\milli\kelvin}.

\begin{figure}[b]
\includegraphics[width=1\columnwidth]{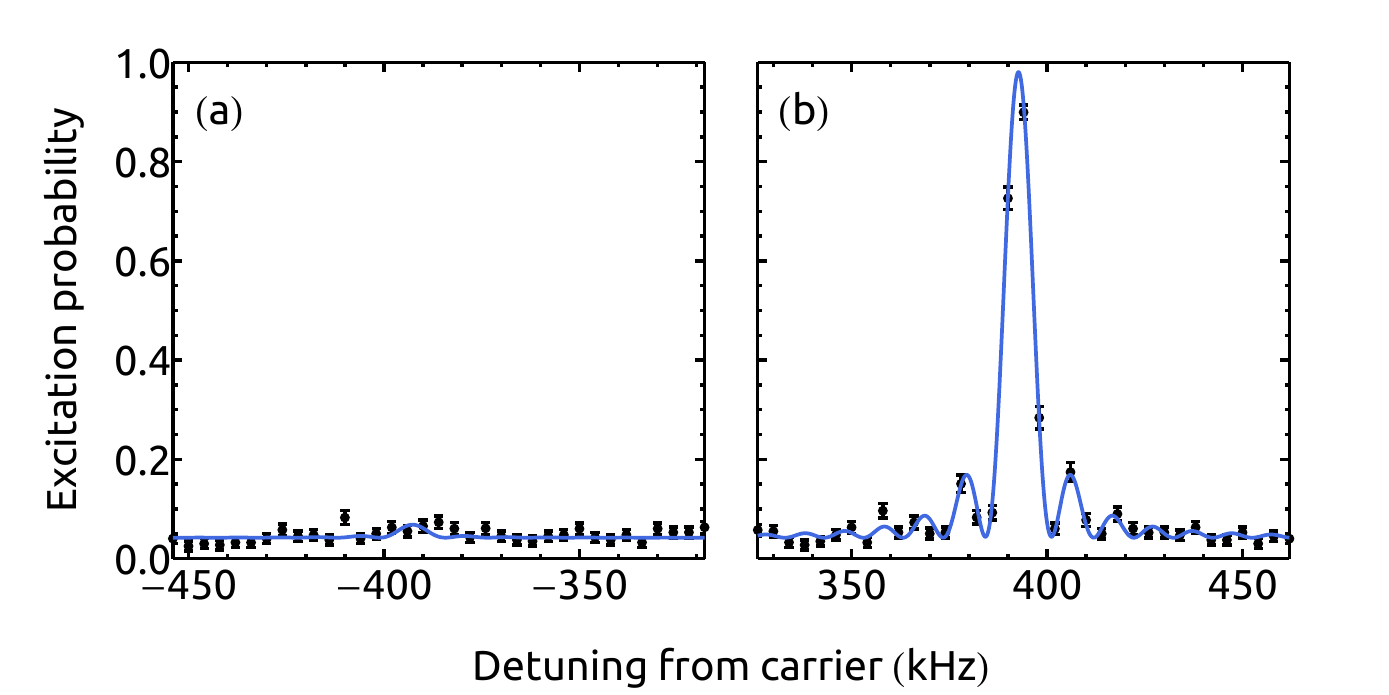}
\caption{(Color online) (a) First red and (b) first blue sidebands after sideband cooling, with a trap frequency of $\omega/2\pi=\SI{389}{\kilo\hertz}$ and a probe time of \SI{100}{\micro\second}. The solid line is a fit to the Rabi dynamics with a constant background, which gives $\bar{n}=\SI{0.02\pm0.01}{}$.}\label{sbcFit}
\end{figure}

Figure \ref{sbcFit} shows the first red and blue sidebands after \SIlist{5;5;10}{\milli\second} of sideband cooling on the first red, second red and first red sidebands sequentially~\footnote{\SI{4}{\milli\second} of cooling is sufficient to reach $\bar{n}\approx0$, but we use a much longer cooling time to ensure minimum $\bar{n}$.}. The red and blue sidebands are fitted simultaneously to Rabi sinc profiles on a constant background~\cite{PhysRevX.2.041014}, with the background amplitude, $\Omega_0$ and $\bar{n}$ as free parameters. The fit shows the average phonon number after sideband cooling to be $\bar{n}=\SI{0.02\pm0.01}{}$, consistent with the theoretical sideband cooling limit of $\bar{n}_{lim}=0.015$ (Eq.~\ref{eq:sbcLimit}).

\begin{figure}
    \includegraphics[width=\columnwidth]{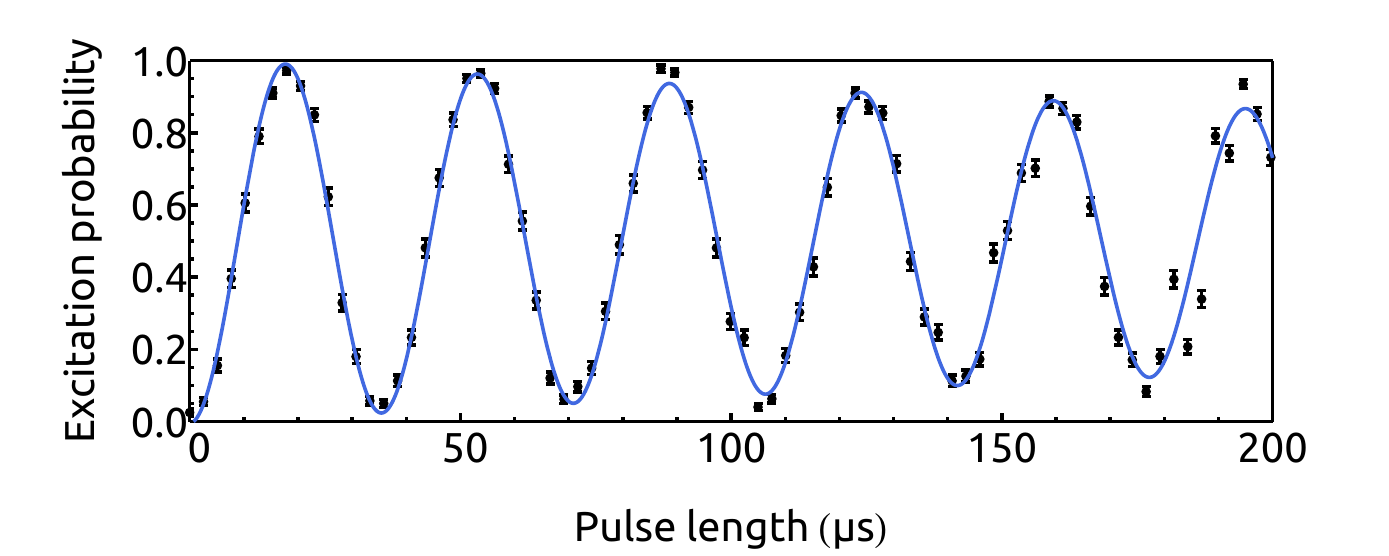}
    \caption{(Color online)
	Rabi oscillation on the carrier after ground state cooling. The Rabi frequency is $\Omega_0 / 2\pi \approx \SI{28}{\kilo\hertz}$ and the overall visibility decays with a coherence time of $\tau\approx\SI{0.7}{\milli\second}$.}\label{Rabi}
\end{figure}

Once cooled to the ground state, dephasing due to thermal effects becomes insignificant, allowing us to perform coherent qubit manipulations. After ground state cooling, we probe the carrier transition and observe damped Rabi oscillations (Figure~\ref{Rabi}). The source of the decoherence cannot be identified from these data so the function describing the damping is unknown. To extract the Rabi frequency and provide an approximate measure of the coherence time, we fit an exponentially decaying sinusoid, which gives $\Omega_0/2\pi=\SI{28}{\kilo\hertz}$ and $\tau\approx\SI{0.7}{\milli\second}$, consistent with our spectroscopically measured linewidth of $\Delta\nu=\SI{0.6\pm0.4}{\kilo\hertz}$~\cite{Mavadia2014}. As the number of oscillations increases, the data increasingly deviate from the theoretical fit. This is predominantly due to intensity noise on the 729-nm probe laser, on a characteristic timescale longer than that taken to record each data point. To reduce this effect we are currently developing a power-noise-eating feedback system for this laser.

The heating rate, $\dot{\bar{n}}$, of the trap is determined by inserting a delay period between ground state cooling and spectroscopy, during which no cooling is applied, and measuring the increase in phonon number. We perform heating rate measurements with delays of \SIlist{0;50;100}{\milli\second}, interleaved in a single experimental pulse sequence. 

\begin{figure}
\includegraphics[width=\columnwidth]{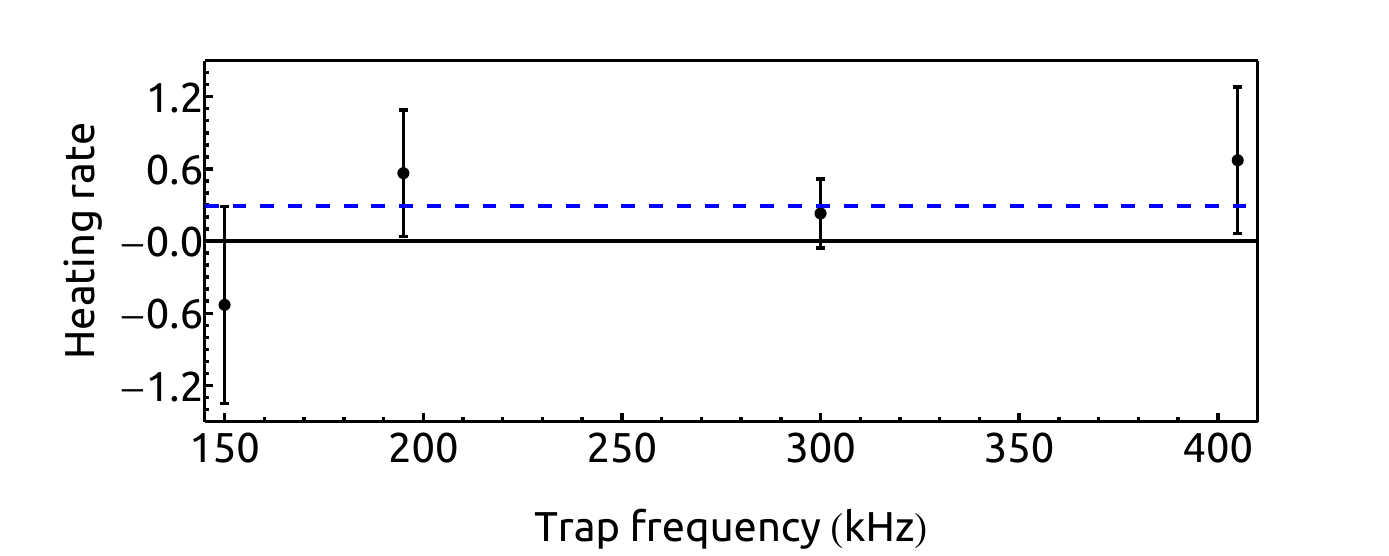}
\caption{(Color online) Heating rate vs axial trapping frequency, $\omega/2\pi$. The data do not indicate a clear frequency scaling; a constant fit gives an average of $\dot{\bar{n}}=\SI{0.3\pm0.2}{\second^{-1}}$.
}\label{heatVfreq}
\end{figure}

To investigate the source of the heating, we repeat this experiment at a range of axial trap frequencies, summarised in Figure~\ref{heatVfreq}. In general, the heating rate for an ion of mass $m$ due to electric field noise with spectral density $S_E(\omega)\propto\omega^{-\alpha}$ is given by $\dot{\bar{n}}=(e^2/4m\hbar\omega)S_E(\omega)$~\cite{Brownnutt2015}. Different sources of electric field noise lead to different characteristic frequency scalings: Johnson noise due to purely resistive elements is independent of frequency ($\alpha=0$); inductive pick-up (e.g. by trap supply cabling) of environmental electromagnetic interference (EMI) leads to $\alpha=-1$ to $1$~\cite{Brownnutt2015} as well as resonances due to local sources; while the patch-potential models developed to describe the $d^{-4}$ ion-electrode distance scaling observed in early experiments predict $\alpha=1$ to $2$~\cite{Hite2013}.

Considering the data in Figure~\ref{heatVfreq}, it is apparent that the heating rate is very low across all trap frequencies, averaging $\dot{\bar{n}}=\SI{0.3\pm0.2}{\second^{-1}}$. The large uncertainties in this dataset prevent us from determining a precise frequency dependence, although it appears very unlikely that an $\omega^{-2}$ or $\omega^{-3}$ scaling is present, as would be expected if fluctuating patch potentials were the dominant heating mechanism. The long delay periods required to measure heating rates of less than one phonon per second reduce the number of repeats it is possible to take for each spectroscopy point, limiting the precision of these measurements. No significant variation of the heating rate has been observed over several months with the current apparatus.

Voltage noise on the electrodes (such as is produced by Johnson noise and inductive pick-up) produces field noise at the ion that scales as $d^{-2}$. When working with larger traps it is likely that such noise sources will eventually become more significant than patch-potential heating. These sources also show weaker scalings with frequency, as suggested by our data. For these reasons we believe that these are the most likely candidates for the source of the inferred electric field noise in our trap, $S_E\approx\SI{5e-16}{\volt^2\m^{-2}\hertz^{-1}}$.

Due to the symmetries of our trap, only differential-mode noise between the trap end-caps or between the compensation electrodes can lead to significant heating of the axial mode. These end caps are individually connected via \SI{3}{\metre} of cabling to a common voltage source. The resistance of this loop of cabling is approximately \SI{1.5}{\ohm}, leading to a Johnson-noise-induced $S_E=4 k_B T R/d^2=\SI{2e-16}{\volt^2\m^{-2}\hertz^{-1}}$ at the ion, independent of frequency. Inductive pickup by this loop is reduced by routing the two cables side by side, minimising the loop area. However even very small linked fluxes can lead to significant field noise. Assuming a typical laboratory EMI noise figure of $F_a=\SI{140}{\decibel}$ at \SI{0.3}{\mega\hertz} with an $\omega^{-5}$ frequency scaling~\cite{Landa2011}, the field noise at the ion due to inductive pickup by a loop of area $A$ would be $S_E\approx A^2\times\SI{e-9}{\volt^2\m^{-6}\hertz^{-1}}$, scaling with $\omega^{-1}$~\cite{Brownnutt2015}. This would exceed our inferred spectral noise for loop areas as small as \SI{5}{\cm^2}. While calculating the exact effects of EMI will require a measurement of the environmental noise in the vicinity of the experiment, it is certainly plausible that a combination of inductive pick-up and Johnson noise could account for the observed heating rate. A simple way to reduce these effects by over two orders of magnitude would be to connect the two end caps at the location of the trap. Similar forms of voltage noise on the compensation electrodes could be minimised by moving the low-pass filters for these electrodes from the remote trap supply to the trap.

\begin{figure}
\includegraphics[width=\columnwidth]{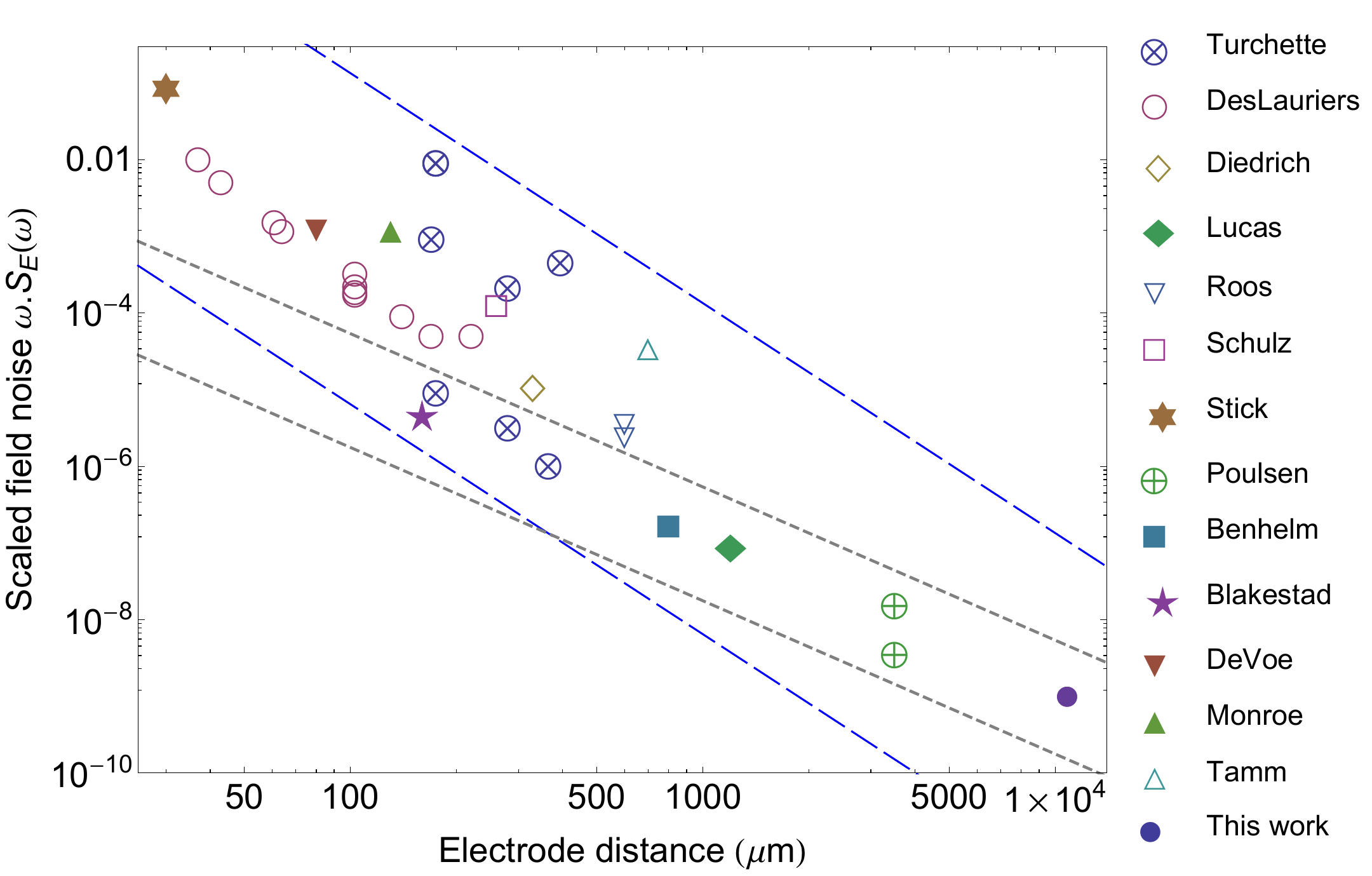}
\caption{(Color online) The reported heating rates of a variety of room temperature, three-dimensional traps, shown as a frequency-scaled, noise spectral density plotted against the distance to the nearest electrode. The result reported in this Letter is shown in the lower right corner. The blue (dashed) and grey (dotted) lines are guides representing $1/d^2$ and $1/d^4$ distance scalings respectively. [Data sources: Turchette \cite{Turchette2000}, DesLauriers \cite{Deslauriers2006}, Diedrich \cite{Diedrich1989}, Lucas ~\cite{Lucas2007}, Roos \cite{Roos1999}, Schulz \cite{Schulz2008}, Stick \cite{Stick2006}, Poulsen \cite{Poulsen2012}, Benhelm \cite{Benhelm2008}, Blakestad \cite{Blakestad2011}, DeVoe \cite{DeVoe2002}, Monroe \cite{Monroe1995}, Tamm \cite{Tamm2000}].}\label{heatingScaling}
\end{figure}

In Figure \ref{heatingScaling} we compare our heating rate to those in a range of other traps, plotted against the distance to the nearest electrode. Here we have taken the usual approach of plotting the phonon heating rate in terms of an inferred scaled noise spectral density, $\omega S_E(\omega)$, that assumes $\alpha=1$. Due to the large uncertainties of our individual measurements of $\dot{\bar{n}}$ and the absence of a clear frequency scaling, we have plotted the average of the noise densities calculated for each frequency. The list of measurements is not exhaustive; from those collated in Ref.~\cite{Amini}, we have included only room temperature traps with three-dimensional electrode structures, and have additionally included the results from Poulsen \emph{et al}~\cite{Poulsen2012} as these were the lowest heating rates previously reported in any trap. Our values of $\omega S_E(\omega)$ are several times lower than those in Ref.~\cite{Poulsen2012}. This is not a particularly surprising result given the large ion-electrode distance in our trap.

We have included guide lines in this figure showing the slope of the expected $d^{-4}$ scaling due to patch potential heating. Our data are approximately consistent with this scaling, although somewhat higher than might be expected. However, if voltage noise sources (e.g. Johnson noise) dominate for large traps, one would expect to see a transition to a $d^{-2}$ scaling at some point, as indicated here by the second pair of guide lines. While there does appear to be some limited evidence of such a transition for $d>\SI{1}{\milli\metre}$, this could well be coincidental. To confirm such a cross-over would require a carefully designed experiment, as variations in the supply electronics and cabling between apparatus would likely outweigh any distance scaling, and indeed would determine the distance at which the transition would occur. It is interesting to note that the results in Ref.~\cite{Poulsen2012} also suggest a heating rate that is constant in frequency, with a greater confidence than that given by our data. This would be consistent with inductive noise due to some forms of EMI.

Achieving 3-dimensional ground-state confinement requires the magnetron mode to be cooled from a typical $\bar{n}$ after Doppler cooling of several thousand. This is further complicated by the non-separability of the radial modes and the negative total energy associated with the magnetron motion. Work on this subject is ongoing, but the process is considerably more challenging than axial cooling.

We have demonstrated the application of resolved sideband laser cooling to an ion in a Penning trap, achieving occupancy of the axial motional ground state with 98\% probability. We have measured the heating rate of our ion trap and found it to be the lowest reported to date, although the uncertainties in the data prevent unambiguous identification of the source of the heating. These results pave the way for an exciting new range of Penning trap experiments in precision measurement~\cite{Smorra2015}, quantum information~\cite{Goodwin2015} and quantum thermodynamics~\cite{Ruiz2014}.

\begin{acknowledgments}
This work was supported by the UK Engineering and Physical Sciences Research Council (Grant EP/D068509/1) and by the European Commission STREP PICC (FP7 2007-2013 Grant 249958). We gratefully acknowledge financial support towards networking activities from COST Action MP 1001 - Ion Traps for Tomorrow's Applications.
\end{acknowledgments}

\bibstyle{plain}

\end{document}